\documentstyle[12pt]{article}
\begin{document}
\vskip 0.1in
\centerline{\Large\bf Cosmology on a Three-Brane}
\vskip .7in
\centerline{Dan N. Vollick}
\centerline{Department of Mathematics and Statistics}
\centerline{and}
\centerline{Department of Physics}
\centerline{Okanagan University College}
\centerline{3333 College Way}
\centerline{Kelowna, B.C.}
\centerline{V1V 1V7}
\vskip .9in
\centerline{\bf\large Abstract}
\vskip .5in
In this paper a general solution is found for a five dimensional
orbifold spacetime that induces a $k=0$ cosmology on a three-brane.
Expressions for the energy density and pressure on the brane in terms
of the brane metric are derived. Given a metric on the
brane it is possible to find five dimensional spacetimes that 
contain the brane. This calculation is carried out for an inflationary
universe and for a metric that corresponds to a radiation 
dominated universe in standard cosmology. It is also shown that
any $k=0$ cosmology can be embedded in a flat five dimensional
orbifold spacetime and the equation of the three-brane surface is derived.
For an inflationary universe it is shown that the surface is the usual
hyperboloid representation of de Sitter space, although it is embedded
in an orbifold spacetime.
\newpage
\section*{Introduction}
Over the last few years there has been a great deal of interest in
the possibility that we live on a three dimensional brane embedded
in a higher dimensional space. Horava and Witten \cite{Ho1} showed that
in the strongly coupled limit $E_8\times E_8$ heterotic string theory
can be viewed as an 11 dimensional theory on the orbifold $R^{10}
\times S^1/Z_2$ with gravitons propagating in the bulk and 
super Yang-Mills fields confined to the two ten-branes that form the 
boundary of the spacetime. Recently a 
novel solution to the hierarchy
problem was proposed \cite{Ar1,An1,Ar2} by considering the 
possibility that $n$ of the compactified 
dimensions are `large'. It was shown that the effective 4 dimensional
Planck mass $M_{pl}$ is related to the 5 dimensional Planck mass $M$ via
$M^2_{pl}\sim M^{2+n}V_{(n)}$, where $V_{(n)}$ is the volume of the
compactified space. Thus, if the extra dimensions are large enough it
is possible to have a small $M$ (even on the order of a TeV) with $M_{pl}
\sim 10^{19}$ Gev. Therefore, the hierarchy problem 
can be solved by reducing the 5
dimensional Planck mass. For $M\sim 1$ Tev they found that $n>2$ to
avoid conflicts with observations if the standard model fields are confined
to a three brane while gravity propagates in the bulk. For $n=2$ 
astrophysical constraints force $M\stackrel{>}{\sim} 100$ Tev 
\cite{Cu1, Ha1} and the size of the dimensions is constrained
to be $\stackrel{<}{\sim} 5\times 10^{-5}$ mm.
    
Two other interesting possibilities were recently suggested by
Randall and Sundrum \cite{Ra1,Ra2}. In their first model they considered
a five dimensional spacetime with the fifth dimension, labeled by $w$, 
compactified on $S^1$ with $-w_c\leq w\leq w_c$ and with
the orbifold symmetry $w\sim -w$. The brane at $w=0$ is a domain wall
with positive tension and the brane at $w=w_c$ is a domain wall with
negative tension. They showed that mass scales on the negative tension
brane can be severly suppressed, leading to a solution of the hierarchy
problem. Of course, this assumes that we live on the negative tension
brane. It was shown by Shiromizu, Maeda, and Sasaki 
\cite{Sh1}
that the effective Einstein field equations on the negative tension brane
involve a negative gravitational constant which means that gravity
would be repulsive instead of attractive. However, they showed that one
does recover the correct Einstein equations in the low energy limit
on the positive tension brane. More recently, it has been shown \cite{Cs2}
that the problem with the negative tension brane may disappear if 
the extra dimension is stabilized by a radion field.
In the second scenario of Randall
and Sundrum we live on the positive tension brane and the negative tension
brane is moved off to infinity. Thus, in this scenario the extra dimension
is infinite in extent. As usual the fields of the standard model
live on the brane and gravity lives in the bulk. They showed that
there is a single gravitational bound state confined to the
brane that corresponds to the graviton. They also showed that,
even though the extra dimension is infinite, the
effective gravitational interaction on the brane is 
that of a four dimensional 
spacetime with some very small corrections.
   
Some aspects of the cosmology of branes embedded in higher dimensional
spacetimes have been examined. Bin\'{e}truy, Deffayet, and Langlois 
(BDL) \cite{Bi1} showed that the scale
factor $a_0$ on a brane with flat spatial sections satisfies
\begin{equation}
\frac{\ddot{a_0}}{a_0}+\frac{\dot{a_0}^2}{a_0^2}=-\frac{k^4_{(5)}}
{36}\rho(\rho+3p)
\label{eq}
\end{equation}
where $k_{(5)}$ is the five dimensional Einstein constant. I have
dropped a term in this equation
proportional to $T_{55}$ by taking the bulk to be empty.
As they pointed out, this equation is similar to one of the equations
from standard cosmology except that the source terms are nonlinear.
This leads to nonstandard cosmological evolution. 
Cs\'{a}ki, Graesser, Kolda, and Terning \cite{Cs1}
then showed that one can get standard evolution at late
times if the brane has a cosmological constant $\Lambda_b$. Note that
$\Lambda_b$
is different than the bulk cosmological constant, as it is confined to
the brane. If $\Lambda_b>>\rho, p$ the right hand side of (\ref{eq})
gives the usual source terms plus nonlinear corrections that will become
small at late times. One remaining difference between this scenario
and the standard one is that equation (\ref{eq}) is second order.
In the standard scenario the evolution is governed by a first order
equation and the initial condition $a(t=0)=a_0$ (if there is an
initial singularity $a_0=0$) is sufficient to determine the
solution for a given equation of state in a $k=0$ universe.
On the other hand, the $k=0$ cosmology on
a brane depends on $\dot{a}(t=0)$ as well. Specific five dimensional
and three-brane cosmologies have also been discussed by various
authors \cite{Ka2,Bi2,Ni1,Lu1,Dv1,Ka1,Be1}.
  
In this paper a general solution is found for a five dimensional
orbifold spacetime that gives a $k=0$ cosmology on a three-brane. The
energy density and pressure on the brane are written in terms of the
metric on the brane. It is shown that it is possible to 
find five dimensional spacetimes that contain a brane with a given metric.
This procedure is carried out for an inflationary universe and for
a metric that correspond to a radiation dominated universe in
standard cosmology. It is also shown that any $k=0$ cosmology can
be embedded in a five dimensional flat 
orbifold spacetime and the embedding
equation is found. For an inflationary universe it is shown that the
surface is the usual hyperboloid representation of de Sitter space,
although it is embedded in an orbifold spacetime.

\section*{Cosmology in the Five Dimensional Space-time}
Here we consider a five dimensional spacetime with $w$ labeling the extra 
dimension. The fifth dimension is taken to be compactified on $S^1$ with
$-w_c\leq w \leq w_c$. An orbifold $S^1/Z_2$ is produced by
identifying $w$ and $-w$, so that the range of $w$ can be taken to be
$0\leq w\leq w_c$. The spacetime contains a three-brane at $w=0$
and at $w=w_c$. These branes will, in general, have non-zero surface
energies and pressures.
We will take the brane at $w=0$ to be the brane we live on 
and the brane at $w=w_c$ will be a hidden brane. 
   
To simplify the form of the metric we can use a Gaussian normal
coordinate system based on the brane at $w=0$. In this coordinate
system $g_{55}=1$ and $g_{5\mu}=0$ ($\mu=0,1,2,3$). Assuming that
the three dimensional spatial sections are flat gives
\begin{equation}
ds^2=-b^2(t,w)dt^2+a^2(t,w)[dx^2+dy^2+dz^2]+dw^2 .
\end{equation}
Since it is always possible to write a two dimensional metric in 
conformally flat form we can take the metric to be
\begin{equation}
ds^2=c(t,w)[dw^2-dt^2]+a^2(t,w)[dx^2+dy^2+dz^2] .
\end{equation}
Transforming to retarded and advanced coordinates $u=t-w$
and $v=t+w$ gives
\begin{equation}
ds^2=-c(u,v)dudv+a^2(u,v)[dx^2+dy^2+dz^2]
\end{equation}
The Einstein field equations $R_{kl}=\lambda g_{kl}$ in the bulk 
give
\begin{equation}
c(\partial^2_u a)-(\partial_u c)(\partial_u a)=0 ,
\label{5}
\end{equation}
\begin{equation}
a(\partial_u\partial_v a)+2(\partial_u a)(\partial_v a)=
-\frac{1}{4}\lambda ca^2 ,
\label{6}
\end{equation}
\begin{equation}
c(\partial^2_v a)-(\partial_v a)(\partial_v c)=0 ,
\label{7}
\end{equation}
and,
\begin{equation}
3c^2(\partial_u\partial_v a)+ac(\partial_u\partial_v c)-a(\partial_u c)
(\partial_v c)=-\frac{1}{2}\lambda ac^3  ,
\end{equation}
where $\lambda=-2/3\Lambda_{(5)}$ and $\Lambda_{(5)}$ is the five dimensional
cosmological constant. First consider the case $\lambda =0$.
The above equations are trivial to integrate. The general solution
for $\partial_u a, \partial_v a\neq 0$ is
\begin{equation}
a(u,v)=[f(u)+g(v)]^{1/3} ,
\label{a}
\end{equation}
and 
\begin{equation}
c(u,v)=c_1f^{'}(u)g^{'}(v)[f(u)+g(v)]^{-2/3},
\label{c}
\end{equation}
where $f(u)$ and $g(v)$ are arbitrary functions of $u$ and $v$
respectively and $c_1$ is an arbitrary constant.
If $\partial_u a=\partial_v a=0$ then
\begin{equation}
a=a_1 \;\;\;\;\;\;\;\;\;\;\;\; c(u,v)=f(u)g(v) .
\label{flat}
\end{equation}
where $a_1$ is a constant. 
Since all of
these solutions correspond to Minkowski spacetime on the brane and in the
bulk they will not be discussed further.
If $\partial_v a=0$ and
$\partial_u a\neq 0$ then
\begin{equation}
a(u)=h(u)\;\;\;\;\;\;\;\;\;\;\;\;\; c(u,v)=c_2h^{'}(u)k(v) .
\label{extra1}
\end{equation}
A similar result holds if $\partial_u a=0$ and
$\partial_v a\neq 0$. 
  
Transforming (\ref{a}) and (\ref{c}) back to $(t,w)$ coordinates 
and imposing the orbifold condition $w \sim -w$ gives
\begin{equation}
ds^2=|\dot{f}(u)\dot{g}(v)|[f(u)+g(v)]^{-2/3}
(dw^2-dt^2)+[f(u)+g(v)]^{2/3}(dx^2+dy^2+dz^2)
\label{metric1}
\end{equation}
where $u=t-|w|$ and $v=t+|w|$ and 
I have absorbed some constants into the definition of $t$ and $w$.
Since the sign of $c_1$ cannot be absorbed the
coefficients $g_{tt}$ and $g_{ww}$ should have a $\pm$ sign. I have forced
the correct spacetime signature by using absolute values instead and by
restricting $f$ and $g$ to be positive. Note that $f$ and $g$
interchange roles as we cross the brane at $w=0$. 
Transforming (\ref{extra1}) back to $(t,w)$ coordinates gives
\begin{equation}
ds^2=|\dot{h}(u)|k(v)(dw^2-dt^2)+h^2(u)(dx^2+dy^2+dz^2) 
\label{metric1b}
\end{equation}
where $k$ has been taken to be positive. This metric corresponds
to a flat five dimensional spacetime since the Riemann tensor
vanishes. However, the geometry induced on the brane will not, in
general, be flat. For notational
simplicity the metric in both cases will be written as
\begin{equation}
ds^2=n^2(t,|w|)[dw^2-dt^2]+a^2(t,|w|)[dx^2+dy^2+dz^2]
\label{metric1c}
\end{equation}
where $n$ and $a$ are given in (\ref{metric1}) or in (\ref{metric1b}).
   
Now consider $\lambda \neq 0$. Here $\lambda$ will be taken to be
positive, as in the Randall-Sundrum model. Eliminating $c(u,v)$ from
(\ref{5}) and (\ref{7}) and integrating gives
\begin{equation}
a(u,v)=H[f(u)+g(v)]
\end{equation}
where $H$, $f(u)$, and $g(v)$ are arbitrary functions. Solving for
$c(u,v)$ from (\ref{5}) and (\ref{7}) gives
\begin{equation}
c(u,v)=\alpha f'(u)g'(v)H'[f(u)+g(v)]
\end{equation}
where $\alpha$ is an arbitrary constant. Substituting $a(u,v)$ and
$c(u,v)$ into (\ref{6}) gives
\begin{equation}
HH''+2(H')^2=-\frac{1}{4}\lambda\alpha H^2H'.
\end{equation}
A first integral of this equation is
\begin{equation}
H^2H'+\frac{1}{16}\lambda\alpha H^4=c_1
\end{equation}
where $c_1$ is an arbitrary constant. The solution is
\begin{equation}
\tan^{-1}[\beta H]-\frac{1}{2}\ln\left[\frac{\beta H+1}{\beta H
-1}\right]=c_2-2(f(u)+g(v))
\end{equation}
where $\beta=(\frac{\alpha\lambda}{16c_1})^{1/4}$ and $c_2$ is an
arbitrary constant. Unfortunately $H$ cannot be found explicitly
from the above solution. To simplify the problem consider the case
$c_1=0$. The solution then simplifies to
\begin{equation}
H=\left[c_2+\frac{\alpha\lambda}{16}(f(u)+g(v))\right]^{-1}.
\end{equation}
Absorbing $\alpha\lambda/16$ and $c_2$ into the functions $f(u)$ and
$g(v)$ gives
\begin{equation}
ds^2=\frac{16}{\lambda}\frac{f'(u)g'(v)}{[f(u)+g(v)]^2}[dt^2-dw^2]
+\frac{1}{[f(u)+g(v)]^2}[dx^2+dy^2+dz^2]
\label{lam}
\end{equation}
where $u=t-|w|$ and $v=t+|w|$ as before. 
\section*{Cosmology on the Brane I: $\lambda=0$}
The four dimensional spacetime on our brane will be described by
the metric
\begin{equation}
ds^2_{(4)}=-|\dot{f}\dot{g}|[f+g]^{-2/3}dt^2+
[f+g]^{2/3}(dx^2+dy^2+dz^2) 
\label{metric2}
\end{equation}
or by
\begin{equation}
ds^2_{(4)}=-|\dot{h}|kdt^2+h^2(dx^2+dy^2+dz^2) 
\label{metric3}
\end{equation}
where $f=f(t), \; g=g(t), \; h=h(t),$ and $k=k(t)$.
The energy-momentum tensor on the brane will have the form
\begin{equation}
T^m_{\;\; n}=\frac{\delta(w)}{n_0}diag(-\rho,p,p,p,0)
\end{equation}
where $n_0=n(t,0)$. It can easily be
calculated using the jump conditions that follow from the Einstein
field equations \cite{Is1}. BDL \cite{Bi1} have
found that
\begin{equation}
\frac{[\partial_wa]}{a_0n_0}=-\frac{k^2_{(5)}}{3}\rho
\label{rho}
\end{equation}
and
\begin{equation}
\frac{[\partial_wn]}{n_0^2}=\frac{k^2_{(5)}}{3}(3p+2\rho)
\label{p}
\end{equation}
where $k_{(5)}$ is the five dimensional Einstein constant, 
and $[\;\;]$ indicates the jump in a quantity, i.e.
\begin{equation}
[\partial_wa]=\partial_wa(0^+)-\partial_wa(0^-) .
\end{equation}
For the metric given in (\ref{metric1}) 
the energy density and pressure on our brane are given by
\begin{equation}
\rho=\frac{2[f+g]^{-2/3}[\dot{f}-\dot{g}]}
{k^2_{(5)}\sqrt{|\dot{f}\dot{g}|}},
\label{e1}
\end{equation}
and
\begin{equation}
p=\frac{[f+g]^{1/3}}{k^2_{(5)}\sqrt{|\dot{f}\dot{g}|}}
\left[\frac{\ddot{g}}{\dot{g}}
-\frac{\ddot{f}}{\dot{f}}\right]-\frac{1}{3}\rho .
\label{p1}
\end{equation}
For the metric given in (\ref{metric1b}) we find that
\begin{equation}
\rho=\frac{6}{k_{(5)}^2h}\sqrt{\frac{|\dot{h}|}{k}}sign(\dot{h})
\label{e2}
\end{equation}
and
\begin{equation}
p=\frac{1}{k_{(5)}^2}[k|\dot{h}|]^{-1/2}\left[\frac{\dot{k}}{k}-
\frac{\ddot{h}}{\dot{h}}\right]-\frac{2}{3}\rho
\label{p2}
\end{equation}
Thus given $f$ and $g$ or $h$ and $k$ we can find the metric and
the energy-momentum tensor on the brane. For example if $f(t)=g(t)$ we find
that
\begin{equation}
ds^2_{(4)}=-d\tau^2+\tau (dx^2+dy^2+dz^2)
\end{equation}
where $d\tau=|\dot{f}|(2f)^{-1/3}dt$. This is the spacetime for a
radiation dominated universe in the standard scenario. However, as can 
be seen from (\ref{e1}) and (\ref{p1}) the pressure and energy density
on the brane vanish! Note that this is consistent with equation (\ref{eq})
found by BDL. It is also important to note that Minkowski space is
also a solution for the spacetime on the brane if $\rho=p=0$. 
Surprisingly, flat spacetime also arises
with $\rho=-3p=(4/k_{(5)}^2)$sign($\dot{h}$) if $f+g$=constant.  
It is important to note that the energy densities and pressures above
arise in the five dimensional Einstein equations. If an observer on
the brane defined $T_{\mu\nu}^{(4)}=k_{(4)}^{-2}G_{\mu\nu}^{(4)}$
the energy densities and pressures obtained would differ from the
above expressions.
Once we have $f(t)$ and $g(t)$, which specify the cosmology on the brane,
we can extend the solution into the bulk by using (\ref{metric2}) 
or (\ref{metric3}). For example if $f+g=$constant, which gives
flat spacetime on the brane,  then
\begin{equation}
ds^2=\dot{f}(t-|w|)^2(dw^2-dt^2)+dx^2+dy^2+dz^2 .
\end{equation}
This is flat five dimensional spacetime.
If $f(t)=g(t)=t$ on the brane
the bulk metric is given by
\begin{equation}
ds^2=-d\tau^2+\tau(dx^2+dy^2+dz^2)+\frac{1}{\tau}dw^2,
\end{equation}
which corresponds to a five dimensional anisotropic Kasner cosmology. 
Of course, for other choices of $f$ and $g$ the bulk spacetime
metric will depend on $w$.
  
For the remainder of this section I will work with metrics of the form
(\ref{metric1b}). As discussed earlier all metrics of this form
correspond to a flat five dimensional spacetime. To see this let
$\bar{u}=h(u)$ and $\bar{v}=k_1(v)=\int k(v)dv$ (for simplicity
I have taken $\dot{h}>0$). 
In these coordinates
the metric is given by
\begin{equation}
ds^2=-d\bar{u}d\bar{v}+\bar{u}^2(dx^2+dy^2+dz^2) .
\end{equation}
The coordinate transformation that takes this to the flat spacetime
metric 
\begin{equation}
ds^2=-d\tilde{u}d\tilde{v}+d\tilde{x}^2+d\tilde{y}^2+
d\tilde{z}^2
\end{equation}
is
\begin{equation}
\begin{array}{cc}
\tilde{u}=\bar{u}\\
   \\
\tilde{v}=\bar{v}+\bar{u}x^2\\
  \\
\tilde{x}^k=\bar{u}x^k
\end{array}
\end{equation}
where $\bar{x}^2=x^kx_k$. This transformation can easily be found
by using the null geodesics as coordinate lines. Thus, the
transformation from $(t,w,x^k)$ to $(\tilde{t},\tilde{w},\tilde{x}^k)$
is given by\begin{equation}
\begin{array}{cc}
\tilde{t}=\frac{1}{2\lambda}[k_1(t+w)+(x^2+\lambda^2)h(t-w)],\\
    \\
\tilde{w}=\frac{1}{2\lambda}[k_1(t+w)+(x^2-\lambda^2)h(t-w)],\\
   \\
\tilde{x}^k=h(t-w)x^k,
\end{array}
\end{equation}
where $\tilde{u}=\frac{1}{\lambda}(\tilde{t}-\tilde{w})$, $\tilde{v}
=\lambda (\tilde{t}+\tilde{w})$, and $\lambda$ is a constant with
dimensions of length. The orbifold in five dimensions
is produced by identifying the points $\tilde{x}^{\mu}(t,x^k,w)$
and $\tilde{x}^{\mu}(t,x^k,-w)$, where $\mu$ labels the five dimensional
spacetime.
The surface of the brane can therefore be
parameterized by
\begin{equation}
\begin{array}{cc}
\tilde{t}=\frac{1}{2\lambda}[k_1(t)+(x^2+\lambda^2)h(t)],\\
   \\
\tilde{w}=\frac{1}{2\lambda}[k_1(t)+(x^2-\lambda^2)h(t)],\\
    \\
\tilde{x}^k=h(t)x^k ,
\end{array}
\end{equation}
where $(t,x^k)$ are the coordinates on the surface.
This surface can also be described by the equation
\begin{equation}
\tilde{t}^2-\tilde{w}^2-\tilde{x}^2-\tilde{y}^2-\tilde{z}^2
=\left(\frac{\tilde{t}-\tilde{w}}{\lambda}\right)k_1
\left[ h^{-1}\left(\frac{\tilde{t}
-\tilde{w}}{\lambda}\right)\right] .
\end{equation}
Now given any metric of the form
\begin{equation}
ds^2_{(4)}=-dt^2+a^2(t)(dx^2+dy^2+dz^2)
\end{equation}
it is easy to see that $h(t)=a(t)$ and $k_1(t)=\int\frac{dt}{
\dot{a}(t)}$. Thus, any $k=0$ cosmology can be embedded in a flat
five dimensional orbifold spacetime. 
For example consider the inflationary cosmology
with $a(t)=e^{Ht}$. The equation of the surface is
\begin{equation}
\tilde{t}^2-\tilde{w}^2-\tilde{x}^2-\tilde{y}^2-\tilde{z}^2=-\frac{1}{H^2} .
\label{ds}
\end{equation}
This is the usual hyperboloid representation
of de Sitter space except that it is embedded in an orbifold spacetime.
   
Finally, we will show that it is possible to find a three brane metric
and the equation of its surface given the energy density on the brane.
Instead of working with the metric in the form (\ref{metric3})
it is convenient to work in a coordinate system in which
the metric takes
the form
\begin{equation}
ds^2_{(4)}=-d\tau^2+a^2(\tau)[dx^2+dy^2+dz^2]
\end{equation}
where $d\tau=\sqrt{k|\dot{h}|}dt$. Equations 
(\ref{e2}) and (\ref{p2}) become
\begin{equation}
\rho=\frac{6}{k_{(5)}^2a}\frac{da}{d\tau} ,
\label{e4}
\end{equation}
and
\begin{equation}
p=-\frac{2}{k_{(5)}^2}\left(\frac{da}{d\tau}\right)^{-1}\frac{d^2a}
{d\tau^2}-\frac{2}{3}\rho .
\label{p4}
\end{equation}
Note that the above are independent of $k$
so that $k$ will remain arbitrary. Inverting (\ref{e4}) gives
\begin{equation}
a(\tau)=h_0\exp\left[\frac{k^2_{(5)}}{6}\int\rho(\tau)d\tau\right] .
\end{equation}
Note that $p$ is fixed once $h$ is known. For example if $\rho=\rho_0$=
constant then $a(\tau)=\exp(H\tau)$ and $p=-\rho$
where $H=k_{(5)}^2\rho_0/6$. The equation of the surface is given in
(\ref{ds}).
\section*{Cosmology on the Brane II: $\lambda > 0$}
From (\ref{lam}) the induced metric on the brane is
\begin{equation}
ds_{(4)}^2=\frac{16}{\lambda}\frac{\dot{f}\dot{g}}{(f+g)^2}dt^2
+\frac{1}{(f+g)^2}(dx^2+dy^2+dz^2)
\end{equation}
and the energy density and pressure on the brane for $f+g>0$ are given by
\begin{equation}
\rho=\frac{3\sqrt{\lambda}}{2k_{(5)}^2}\left[\frac{\dot{g}-\dot{f}}
{\sqrt{-\dot{f}\dot{g}}}\right]
\end{equation}
and
\begin{equation}
P=\frac{\sqrt{\lambda}}{2k_{(5)}^2\sqrt{-\dot{f}\dot{g}}}\left[
3(\dot{f}-\dot{g})+\frac{1}{2}(f+g)\left(\frac{\ddot{g}}{\dot{g}}
-\frac{\ddot{f}}{\dot{f}}\right)\right] .
\end{equation}
Note that $\dot{f}\dot{g}<0$ for the metric on the brane to have the
correct signature and for the energy density and pressure to be real.
  
As an example consider the spacetime with $f(t)=1/t$, $g(t)=t$, and $t>0$. 
The metric on the brane is
\begin{equation}
ds_{(4)}^2=-d\tau^2+\frac{1}{4}\sin^2\left(\frac{\sqrt{\lambda}\tau}
{2}\right)[dx^2+dy^2+dz^2]
\end{equation}
where $\tau=\frac{4}{\sqrt{\lambda}}\tan^{-1}(t)$. This describes a
cosmology in which the universe initially expands and then collapses.
The energy density and pressure on the brane are given by
\begin{equation}
\rho=\frac{3\sqrt{\lambda}}{2k_{(5)}^2}\left[\sin\left(\frac{\sqrt{\lambda}
}{2}\tau\right)\right]^{-1}
\end{equation}
and
\begin{equation}
P=-\frac{2}{3}\rho.
\end{equation}
It is important to note that the energy density and pressure found above
arise in the five dimensional Einstein equations. If an observer on the
brane defined $T_{\mu\nu}^{(4)}=k_{(4)}^{-2}G_{\mu\nu}^{(4)}$ 
different results would be
obtained.
\section*{Conclusion}
A general solution was found for a five dimensional orbifold
spacetime that induces a $k=0$ cosmology on a three-brane. Expressions
for the energy density and pressure on the brane where found in terms
of the metric on the brane. 
It was shown that it is possible to find five
dimensional spacetimes that contain the brane given the brane metric.
This procedure was carried out for scale factors
$a(\tau)=e^{H\tau}$ and $a(\tau)=\tau^{1/2}$. It was also shown that any $k=0$
cosmology can be embedded in a flat five dimensional orbifold spacetime
and the embedding equation was found. For an inflationary universe it
was shown that the surface is the usual hyperboloid representation
of de Sitter space, although it is embedded in an orbifold spacetime.

\end{document}